\documentclass[conference]{IEEEtran}
\IEEEoverridecommandlockouts

\usepackage{graphicx, caption, subcaption}

\usepackage[utf8]{inputenc}
\usepackage{amsmath,amsfonts}
\usepackage{algorithmic}
\usepackage{textcomp}
\usepackage[inline]{enumitem}
\usepackage{url}

\usepackage[T1]{fontenc}
\usepackage[utf8]{inputenc}
\usepackage{nicefrac}
\usepackage{siunitx}  
\usepackage{flushend}
\usepackage{hyperref}

\usepackage{amsmath,amsthm,amssymb,amsfonts}
\usepackage[table,xcdraw]{xcolor}
\definecolor{Gray}{gray}{0.85}
\usepackage[scientific-notation=false]{siunitx}

\newcommand{\toolname}{PatchScope}

\usepackage{booktabs}
\usepackage{rotating}
\usepackage{multirow}
\usepackage{float}
\floatstyle{plaintop}
\restylefloat{table}
\newcolumntype{a}{>{\columncolor{Gray}}c}
\SetEnumitemKey{midsep}{topsep=3pt, partopsep=0pt}
\usepackage{todonotes}
\newif\iftodos
\todostrue

\newcommand{\DCVE}{$D_{\text{\itshape CVE}}$}
\newcommand{\DCRAWL}{$D_{\text{\itshape CRAWL}}$}
\newcommand{\DBIP}{$D_{\text{\itshape BIP}}$}

\iftodos
\newcommand{\PP}[1]{\todo[inline]{#1}}
\newcommand{\JN}[1]{\todo[inline,color=green!40]{JN: #1}}
\newcommand{\jn}[1]{\todo[color=green!40,size=\tiny]{JN: #1}}
\else
\newcommand{\PP}[1]{}
\newcommand{\JN}[1]{}
\newcommand{\jn}[1]{}
\fi

\newcommand{\dsname}{HaPy-Bug}

\def\BibTeX{{\rm B\kern-.05em{\sc i\kern-.025em b}\kern-.08em
    T\kern-.1667em\lower.7ex\hbox{E}\kern-.125emX}}
\begin{document}

\title{\dsname\ -- Human Annotated Python Bug Resolution Dataset}

\author{
  \IEEEauthorblockN{
    Piotr Przymus\IEEEauthorrefmark{1},
    Mikołaj Fejzer\IEEEauthorrefmark{2},
    Jakub Narębski\IEEEauthorrefmark{3},
    Radosław Woźniak\IEEEauthorrefmark{4}, \\
    Łukasz Halada\IEEEauthorrefmark{5}, 
    Aleksander Kazecki\IEEEauthorrefmark{6},
    Mykhailo Molchanov\IEEEauthorrefmark{7} and 
    Krzysztof Stencel\IEEEauthorrefmark{8} 
  }
  \textit{Nicolaus Copernicus University},\quad
  \IEEEauthorrefmark{5}\textit{University of Wrocław},\quad
  \IEEEauthorrefmark{8}\textit{University of Warsaw},\quad
  \IEEEauthorrefmark{7}\textit{Kyiv Polytechnic Institute}\\
  Toruń, Poland \hspace{2.5cm}
  Wrocław, Poland \hspace{1.5cm}
  Warsaw, Poland \hspace{2cm}
  Kyiv, Ukraine\\
  Email:
  \IEEEauthorrefmark{1}piotr.przymus,
  \IEEEauthorrefmark{2}mfejzer,
  \IEEEauthorrefmark{3}jakub.narebski,
  \IEEEauthorrefmark{4}rwozniak,
  \IEEEauthorrefmark{6}olekkazecki@mat.umk.pl,\\
  \IEEEauthorrefmark{5}lukasz.halada@cs.uni.wroc.pl,
  \IEEEauthorrefmark{7}m.molchanov.ia12@kpi.ua,
  \IEEEauthorrefmark{8}stencel@mimuw.edu.pl
}

\IEEEtitleabstractindextext{
\begin{abstract}
We present \dsname{}, a curated dataset of 793 Python source code commits associated with bug fixes, with each line of code annotated by three domain experts. The annotations offer insights into the purpose of modified files, changes at the line level, and reviewers' confidence levels. We analyze \dsname{} to examine the distribution of file purposes, types of modifications, and tangled changes. Additionally, we explore its potential applications in bug tracking, the analysis of bug-fixing practices, and the development of repository analysis tools. \dsname{} serves as a valuable resource for advancing research in software maintenance and security.
\end{abstract}

}

\maketitle
\IEEEdisplaynontitleabstractindextext


\maketitle

\makeatletter
\newcommand\footnoteref[1]{\protected@xdef\@thefnmark{\ref{#1}}\@footnotemark}
\makeatother

\section{Introduction}\label{introduction}
We present \dsname, a curated and annotated dataset\footnote{\textit{Supported by the National Science Centre, Poland, project no. 2022/06/X/ST6/00806.}} consisting of 793 Python source code commits linked to bug fixes. Each commit has been manually annotated by three domain experts, capturing the purpose of modified files, bug types, and detailed line-level changes.

Our annotation protocol builds upon the approach by Herbold et al.~\cite{herboldFinegrainedDataSet2022}, introducing significant modifications to improve usability and accuracy. While Herbold et al. used non-trivial automated labels for refactoring, we adopt a different strategy: we provide preliminary annotations based on lexical analysis and file paths, requiring users to refine labels where they find inaccuracies. This contrasts with Herbold et al.'s approach, which simplified rare cases (e.g., automation for refactorings) but left common, often trivial tasks to manual labor prone to error. Our approach shifts the focus to non-trivial annotations, allowing users to correct only where automatic labels fall short, reducing repetitive work and enhancing efficiency. We also introduced new joint categories to cover edge cases.

\dsname{} supports a variety of applications, including bug localization, automated code annotation, enhancing code review workflows, and some code oriented aspects of Large Language Models (LLM).
It has already been utilized in studies on bug localization~\cite{przymusProjectApplicationsGrayscale} and improving automated code annotation~\cite{piotr_przymus_patchscope_2024}.
\dsname{} is a collection of artifacts designed for various types of ``bugs.'' 
We developed \dsname{} as a heterogeneous resource, aiming to encompass a broad spectrum of the otherwise ambiguous concept of a ``bug''~\cite{DBLP:journals/cacm/WidderG24}.

This paper is organized as follows: (1) we describe the dataset collection process; (2) we detail the annotation methodology; (3) we analyze the dataset; (4) we explore its applications; and (5) we discuss its limitations.

The resulting {\bf dataset}, along with {\bf custom tools} and {\bf scripts} developed for this study, is publicly available at \url{https://doi.org/10.6084/m9.figshare.24448663}.

\section{Data Collection and Dataset Structure}\label{data-collection}

\begin{figure*}[h]
  \centering
  \includegraphics[width=0.7\textwidth, trim={0 9.25cm 0 1.5cm},clip]{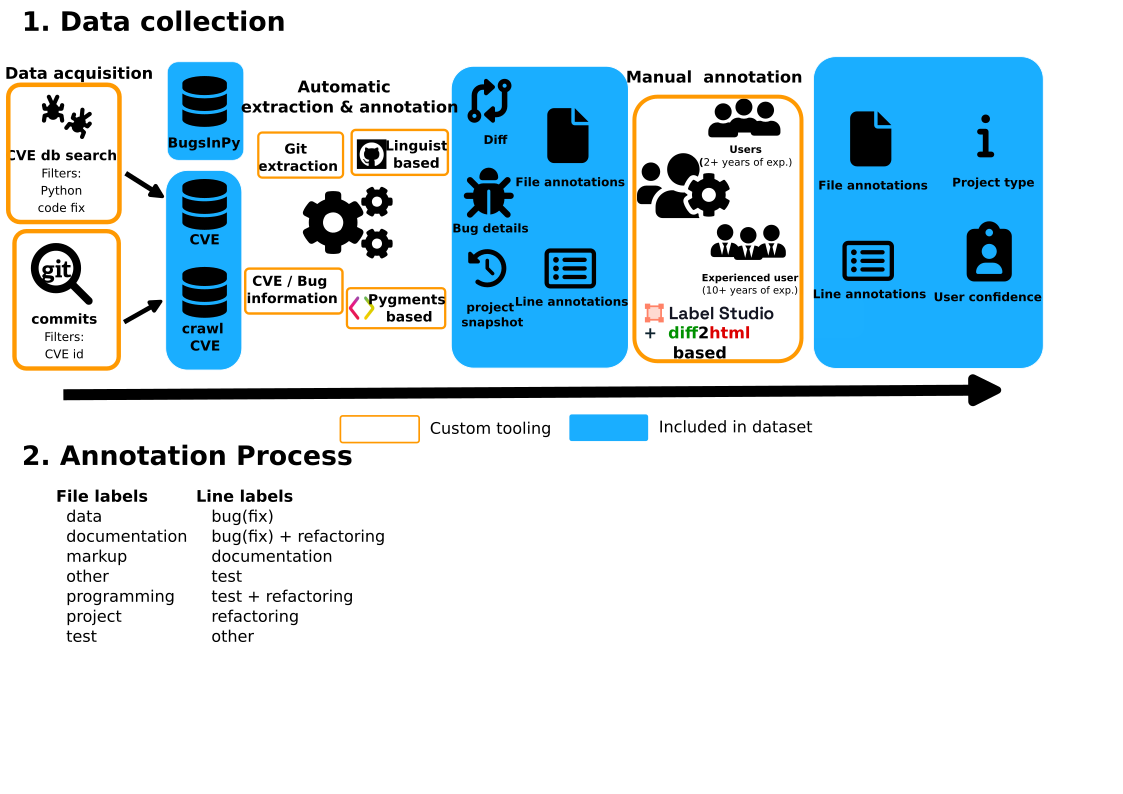}
  \caption{The process of dataset creation}\label{fig:data-process}
\end{figure*}

\dsname\ comprises annotated diff files with fixes of Python-related bugs from three sources.
None of them had previously been annotated by developers at the line level.
In total, \dsname\ encompasses 793 bugs spanning from 2006 to 2022, with code changes distributed across \num{2742} files and covers \num{67963} lines of code (see Tab.~\ref{table:data}).
We provide detailed annotations and comprehensive information for each of the bugs (diff file, file annotation, bug annotations and in case of CVE based bugs --- detailed CVE information and references).
Annotation process will be described in the next section.
The data gathering process is shown in Fig.~\ref{fig:data-process} and described below.

\DBIP{}: \textbf{BugsInPy} is an extension of a dataset of 496 real bugs proposed in~\cite{DBLP:conf/sigsoft/WidyasariSLQPTT20}.
It focuses on isolated bugs in source code and excludes issues related to configurations, build scripts, documentation, and test cases.
Bugs in this set have to be reproducible, i.e.\ at least one test case from the fixed version must fail when executed on the faulty version.
This dataset was manually annotated and incorporated into our dataset.

\begin{table*}[htb]
\centering
\caption{Dataset characteristics, line annotations structure, and user annotation consensus. Line annotations: B:  bug(fix), B+R: bug(fix) + refactoring, D: documentation, O: other, R: refactoring, T: tests, T+R: tests + refactoring.}
\label{table:data}
\setlength{\tabcolsep}{5pt}
\begin{tabular}{@{}lrrrrraaaaaaall@{}}
\toprule
         &  \multicolumn{5}{c} {\bfseries Statistics} & \multicolumn{7} {a} {\bfseries Line annotations} &  \multicolumn{2} {c} {\bfseries Consensus}  \\
 Dataset & \makebox[0.7cm][r]{Repos} & Bugs &  Files &  Lines & \multicolumn{1}{c}{Dates} &  B &  B+R &  D &  O &  R &  T &  T+R &$=3$ &   $\geq 2$ \\

\midrule
\DBIP{}   & 17 &   496 &  1382 &  20221 & 2012--2020 &    24.7\% &                    2.6\% &          13.2\% &   0.5\% &         1.9\% & 55.2\% &                1.9\% & 84.2\% &  97.8\% \\
\DCRAWL{} & 41 &  145 &    759 &  26102 & 2006--2022 &     26.1\% &                    2.1\% &          29.2\% &   1.2\% &         1.5\% & 38.8\% &                1.2\% & 84.2\% &  98.7\% \\
\DCVE{}   & 91 &  152 &    601 &  21640 & 2011--2022 &     29.6\% &                    2.2\% &          14.2\% &   1.2\% &         9.5\% & 42.0\% &                1.3\% & 81.5\% &  98.5\% \\
\bottomrule
\end{tabular}
\end{table*}

\DCVE{}: \textbf{Python CVE} and \DCRAWL{}: \textbf{Crawled Python CVE} are new custom made, specialized collections of bugs sourced from the CVE DB~\cite{cve-db} and git repositories.
\DCVE{} includes bugs identified through a full-text search of the CVE Database~\cite{cve-db}. Only bugs with fixes involving at least one Python file and explicitly referenced in the CVE description were included.
\DCRAWL{} is a subset derived from scanning the top 100 most popular Python repositories on GitHub for commits containing a CVE ID pattern.
Each identified bug was cross-referenced with the CVE Database~\cite{cve-db}, and the associated CVE summaries were used as bug descriptions.

\section{Annotation Process}\label{annotation-process}

We followed a systematic protocol to annotate Python bug-fix commits extending the approach proposed by Herbold et al.~\cite{herboldFinegrainedDataSet2022}.
Fig.~\ref{fig:data-process} provides an overview of the entire process.
This process included classifying files, individual changed lines, project types, and assessing reviewer confidence.
To assist annotators, we automatically suggested three types of line labels—\emph{bug(fix)}, \emph{documentation}, and \emph{test}—based on file paths and syntax highlighting provided by Pygments~\cite{WelcomePygments}.

The \emph{manual annotation process} was carried out by six annotators, each with at least four years of Python programming experience.
Three annotators had over a decade of expertise.
To ensure robust evaluation, the annotation framework assigned each bug-fix resolution to three annotators, including at least one experienced reviewer.
On average, regular annotators reviewed around 600 bugs, while experienced annotators handled approximately 200.

We created an annotation protocol and prepared 10 model annotations as a reference standard\footnote{See \texttt{code/Documentation/} at \url{https://github.com/ncusi/HaPy-Bug}}. 
These were finalized through consensus among experienced annotators.

The next stage involved annotating files and lines, starting with preliminary annotations generated by custom tools.
GitHub Linguist~\cite{Linguist2023} was used to identify file types in bug-related commits.
File purposes were inferred based on types and paths. 
Each changed file was preassigned to a purpose category: ``programming'' (non test source code), ``tests'' (test source code and related files), ``project'' (build tools and dependency management), ``documentation'' (text documentation without markup files), ``markup''(markup files like html), ``data''(non test data files) or ``other'' (rest of the files).
Lines within files were automatically labeled as ``bug(fix)'', ``test'', or ``documentation''.

This process categorizes lines as follows: lines containing only comments and all lines within documentation files are marked as "documentation"; all remaining lines in test files are labeled as "test"; and all other lines, including those in non-programming files, are classified as "bug(fix)."

Annotators reviewed error descriptions sourced from commit and/or CVE metadata, and updated labels as necessary.
All participants annotated an equal amount of bugs from each dataset, with each subset randomly selected.
They could adjust file purposes and apply extended line annotations.
Manual line annotations included all base categories plus additional ones:
\begin{itemize}[nosep]
\item ``refactoring'' for changes improving code readability or structure without altering behavior,
\item tangled categories such as ``bug(fix) + refactoring'' and ``test + refactoring'' for lines combining base categories with refactoring.
\item ``other'' for new functionality, non-bug fixes, etc.,
\end{itemize}

The annotation process lasted two months.
Annotators also specified the type of project, indicated confidence levels, and documented any issues encountered.
Weekly meetings were held to discuss problematic bug reports, resolve ambiguities, and establish consensus on handling specific cases.
One of most common issues present on those meetings was related to the scope of tangled changes for specific bugs.

\section{Data Analysis}\label{data-analysis}

    \begin{figure*}[htb]
    \begin{subfigure}[b]{0.47\textwidth}
      \centering
        \includegraphics[width=0.90\textwidth, trim={0 10.5cm 0 0},clip]{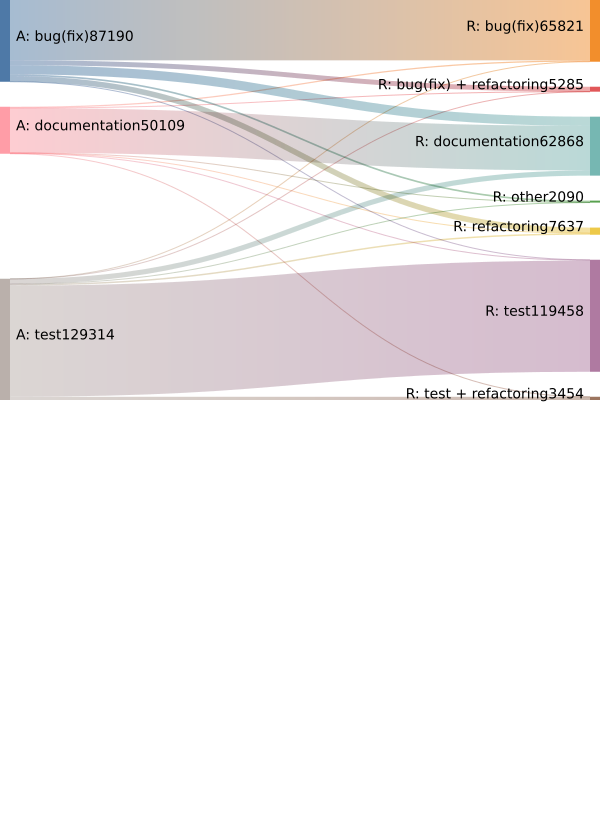}
        \caption{Adjustments performed by individual human annotators.\\ (A)utomated annotation, (R)evised manual annotation.}\label{figadjustments}
    \end{subfigure}
    \begin{subfigure}[b]{0.47\textwidth}
      \centering
        \includegraphics[width=0.90\textwidth, trim={0 0 0 0.3cm}]{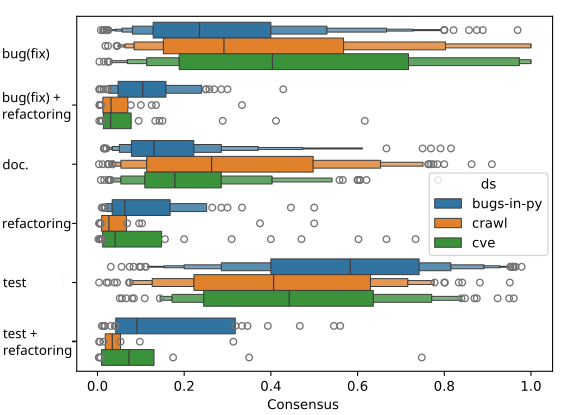}
        \caption{Distribution number of lines types divided by changes in a fix. }\label{figlabels}
    \end{subfigure}
\\
    \begin{subfigure}{1.0\textwidth}
      \centering
      \includegraphics[width=0.8\textwidth, trim={0 0 0 0.3cm},clip]{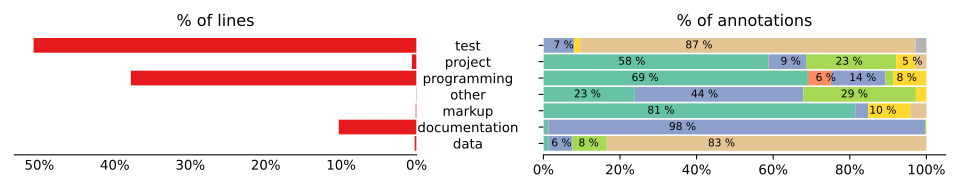}
      \includegraphics[width=0.8\textwidth]{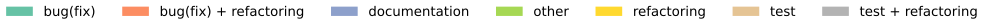}
      \caption{Tangled changes: Percentage (right) of lines by annotated file type and (left) breakdown of line types by file type.}\label{figannotations}
    \end{subfigure}

    \caption{Dataset analysis}

\end{figure*}

In this section, we assess the quality of annotations and automation.
We also examine the characteristics of the dataset, comparing them to those reported in other relevant studies.

\subsubsection{Quality of manual annotations}%
\label{ssub:Quality of manual annotation}

We began by analyzing the multi-rater agreement between reviewers. Using Fleiss' kappa~\cite{fleiss1971measuring} on combined participant annotations across all datasets, we obtained an overall value of $\kappa = 0.83$.
The agreement levels for individual subsets were as follows: \DCVE{} ($\kappa = 0.82$), \DCRAWL{} ($\kappa = 0.84$), and \DBIP{} ($\kappa = 0.82$).
According to Fleiss' kappa interpretation~\cite{fleiss1971measuring}, these values indicate  \textit{``almost perfect agreement''} (highest category).

Next, we analyzed the consensus among annotators.
A line label was considered a full consensus, 
if all three participants selected the same label.
The percentage of full consensus labels varied across different subsets of bugs. 
We observed (see Tab.~\ref{table:data}) that full consensus was achieved for at least $81.5\%$ of labels in the \DCVE{} subset and at least $84.2\%$ in the remaining subsets.
If we consider consensus when at least two reviewers agree the consensus rate increases significantly to $\geq{}98.7\%$.
The above findings demonstrate a high level of confidence in the quality of the annotations.
Notably, the lowest consensus rate occurred for lines labeled as "bug(fix)" by the automated process, likely due to the inherent complexity in identifying bug fixes.

\subsubsection{Quality of automatic suggestions}%
\label{ssub:Quality of automatic suggestions}
We assessed the degree of modifications made by annotators to the automatically generated labels (see Fig.~\ref{figadjustments}).
The latter were based on simple syntax rules and did not account for refactoring suggestions. 
As expected, some corrections were required, especially for documentation due to lack of context in diff files.

We also measured the median percentage of lines within a single bug fix that did not require corrections by reviewers after automatic annotation.
Median percentage of lines not requiring fixes per subset (computed with 95\% Confidence Intervals denoted further as 95\% CI.): \DCVE\ and \DCRAWL\ ($95\%$ CI: $[94\%, 98\%]$), \DBIP\ ($95\%$ CI: $[100\%, 100\%]$).
These findings show that only a small proportion of lines in single bugfix ($\leq 6\%$) needed manual intervention. 
This makes fully automated annotations a viable option for specific applications.

\subsubsection{Dataset characteristics}%
\label{ssub:subsubsection name}
We assess whether bug fixes vary across subsets by testing two hypotheses: 

\textbf{H1:} The proportions of label types differ between subsets.
As we deliberately used 3 different selection criteria for bug fixes, we have unique opportunity to compare how code fixes are constructed depending on the type of fix.
We use $\chi^2$ test of independence of observed frequencies for subsets \DCVE{}, \DCRAWL{} and \DBIP{} with the null hypothesis that the proportions of label types are equivalent between subsets.
We rejected that hypothesis with $p$-value $\leq 0.001$ for all subsets and when compared in pairs. 
Therefore we concluded that they followed different distribution of labels.

For a more detailed view see Tab.~\ref{table:data} and Fig.~\ref{figlabels}, where we can observe that the bug selection criteria significantly influence the dataset's characteristics.
\DBIP{} concentrates on isolated bug fixes with tests for given change.
\DCVE{} represents fixes for bugs that were reported in CVE (a higher emphasis on the code)
\DCRAWL{} is dominated with fixes in dependencies (a high count of changes in documentation and project files).

\textbf{H2}: Less than $40\%$ of changed lines in bug fixing commits contribute to the bug fix.
\noindent This hypothesis derived from~\cite{herboldFinegrainedDataSet2022} and based on~\cite{mills_are_2018} was not rejected in~\cite{herboldFinegrainedDataSet2022}. 
Given the diversity in our dataset subsets, we tested this hypothesis on our data.

In each dataset less than $33\%$ of lines are linked to bug fixes (Tab.~\ref{table:data}), with the majority of changes occurring in production code (Fig.\ref{figannotations}). 
However, when considering individual bug fixes, an analysis of the median proportion of bug fixes per commit (see Fig.~\ref{figlabels}) reveals that \textbf{H2} is rejected for \DCVE{} ($95\%$ CI $[34.0\%, 46.7\%]$).
In contrast, \textbf{H2} holds for \DCRAWL{} ($95\%$ CI $[22.2\%, 37.6\%]$) and \DBIP{} ($95\%$ CI $[22.2\%, 25.3\%]$).
Thus \textbf{H2}'s validity depends on the bug report selection criteria.

\section{Applications}\label{applications}
This project aimed to develop a dataset for precise bug localization at the file and line levels. 
The detailed annotations in \dsname{} improve bug tracking by identifying relevant modified lines.
Human-evaluated annotations aided prioritization and confirmed the viability of automation. 
This use case was the primary motivation and was further explored in \cite{przymusProjectApplicationsGrayscale}.

This dataset was also used to fine-tune and validate \toolname~\cite{piotr_przymus_patchscope_2024}, enhancing automated annotations. 
It helped develop annotation rules for Python projects.
The goal of \toolname{} is to streamline Git repository analysis and provide insights into developer contributions.

Such datasets also enable a deeper understanding of developer practices related to structuring bug fixes.
Studies like Sobreira et al.~\cite{sobreira_dissection_2018} and Herbold et al.~\cite{herboldFinegrainedDataSet2022} use these datasets to analyze common bug-fixing practices, such as commit tangling, the extent of production code changes, and the role of refactoring.
\dsname{} can serve a similar purpose, thanks to the inclusion of line level annotations.

\dsname{} could be used to fine tuning and testing LLMs’ code comprehension, completion and automated program repair under selective code alterations.
This approach evaluates LLM ability to infer missing information and their resilience to adversarial attacks~\cite{DBLP:journals/corr/abs-2312-04730}.

\section{Threats to Validity}\label{limitations}

\indent %
\textbf{Impact of automation}: 
Annotations could be biased through the use of the automation. 
While we acknowledge this risk, we note that our labeling mechanism was essentially based on lexical analysis and rules derived from file types.

\textbf{Bug Scope and Specialization}:
\dsname\ focuses on real bugs within Python source code, adhering to specific selection criteria such as isolated bugs.
This specialization potentially limits its applicability in wider contexts.

\textbf{Inter-Annotator Variability and Annotation Complexity}:
Despite the expertise of domain expert involved, variations in annotations may still occur across different projects.
The complexity of the annotation process introduces the potential for human errors and inconsistencies.

\textbf{Ethical Considerations}: 
We refrain from handling any sensitive contributor or annotator data.
We do not engage in automated judgments to attribute characteristics to individuals.
We asked annotators during each weekly meeting if they have any problems and advised them to limit daily annotation to not exceed 20 bugs per day.

\section{Related Work}\label{related-work}

A number of interesting datasets containing code related to bug fixes exist, some of which include line-level annotations.

\dsname\ expands BugsInPy \cite{DBLP:conf/sigsoft/WidyasariSLQPTT20} by offering human annotations for all collected modifications and by adding two additional, custom bug datasets.

An approach similar to ours is used by Herbold et al.~\cite{herboldFinegrainedDataSet2022} who designed a crowd sourcing procedure to annotate individual lines in Java based projects.
Compared to~\cite{herboldFinegrainedDataSet2022} we added new annotation labels, removed white space annotations, and changed the way suggestions for annotators were provided.
We significantly simplified the process by automatically marking all lines based on straightforward and deterministic lexical analysis (see Sec.~\ref{annotation-process}).
Thus annotators could focus solely on the described bug and address non-trivial markings only.

GitBug~\cite{silva2024gitbug} contains 199 Java bugs extracted from 55 open-source repositories.
Each bug was replicated on a local environment to exclude interference by external parties.
Additionally tests cases were run multiple times to ensure non-flakiness of bug reproduction.
BugsPHP~\cite{pramod2024bugsphp} is the first large dataset targeting PHP language.
The authors collected 600000 bug-fixing commits by crawling 5000 GitHub repositories. 
While both GitBug and BugsPHP are designed for automated program repair, they do not contain line level annotations.

BUGSJS~\cite{DBLP:journals/stvr/GyimesiVSMBFM21} is composed of 453 manually verified JavaScript bugs with corresponding fixes, obtained from 10 popular GitHub JavaScript projects.
Its authors built a taxonomy of bugs by manually labeling each bug. 
While offering interesting insight into the JavaScript ecosystem, BUGSJS does not offer line level annotations per each bug.

Two other interesting datasets PySStuBs
\cite{DBLP:conf/msr/KamienskiPBH21} and TSSB-3M
\cite{DBLP:conf/msr/RichterW22} are devoted to the scrutiny of single
line/statement bugs. \dsname\ concentrates more on complex bugs.

The Defectors \cite{DBLP:conf/msr/MahbubSR23} dataset contains a significant number of source code files.
It contains automatic annotations, while \dsname\ contains both manual and automatic annotations.

To summarize, \dsname\ is a versatile high quality dataset that stands out due to its substantial size.
It encompasses both file- and line-level manual annotations and offers a comprehensive understanding of bug resolution processes. 

\section{Conclusion}\label{conclusion}
We presented \dsname{}, a curated and annotated bug dataset, enriched through a comprehensive annotation protocol with custom tooling used in the process.
It can serve as a valuable asset for the software development community.
With a focus on Python source commits related to bug fixes, this dataset provides deep insights into the world of bug resolution.
The data offers not only an analysis of file purposes and modified line annotations, but also the level of bugfix tangling.

The potential applications of \dsname{} are diverse, spanning bug localization, fine tuning tools for analysis of Git repositories, or research on bugfixing practices.
Moreover, \dsname{} has some potential in research on LLMs in terms of code completion and comprehension.
\newpage

\bibliographystyle{IEEEtran}
\bibliography{datasets.bib}

\begin{thebibliography}{10}
\providecommand{\url}[1]{#1}
\csname url@samestyle\endcsname
\providecommand{\newblock}{\relax}
\providecommand{\bibinfo}[2]{#2}
\providecommand{\BIBentrySTDinterwordspacing}{\spaceskip=0pt\relax}
\providecommand{\BIBentryALTinterwordstretchfactor}{4}
\providecommand{\BIBentryALTinterwordspacing}{\spaceskip=\fontdimen2\font plus
\BIBentryALTinterwordstretchfactor\fontdimen3\font minus \fontdimen4\font\relax}
\providecommand{\BIBforeignlanguage}[2]{{%
\expandafter\ifx\csname l@#1\endcsname\relax
\typeout{** WARNING: IEEEtran.bst: No hyphenation pattern has been}%
\typeout{** loaded for the language `#1'. Using the pattern for}%
\typeout{** the default language instead.}%
\else
\language=\csname l@#1\endcsname
\fi
#2}}
\providecommand{\BIBdecl}{\relax}
\BIBdecl

\bibitem{herboldFinegrainedDataSet2022}
S.~Herbold, A.~Trautsch, B.~Ledel, A.~Aghamohammadi, T.~A. Ghaleb, K.~K. Chahal, T.~Bossenmaier, B.~Nagaria, P.~Makedonski, M.~N. Ahmadabadi, K.~Szabados, H.~Spieker, M.~Madeja, N.~Hoy, V.~Lenarduzzi, S.~Wang, G.~{Rodr{\'i}guez-P{\'e}rez}, R.~{Colomo-Palacios}, R.~Verdecchia, P.~Singh, Y.~Qin, D.~Chakroborti, W.~Davis, V.~Walunj, H.~Wu, D.~Marcilio, O.~Alam, A.~Aldaeej, I.~Amit, B.~Turhan, S.~Eismann, A.-K. Wickert, I.~Malavolta, M.~Sul{\'i}r, F.~Fard, A.~Z. Henley, S.~Kourtzanidis, E.~Tuzun, C.~Treude, S.~M. Shamasbi, I.~Pashchenko, M.~Wyrich, J.~Davis, A.~Serebrenik, E.~Albrecht, E.~U. Aktas, D.~Str{\"u}ber, and J.~Erbel, ``A fine-grained data set and analysis of tangling in bug fixing commits,'' \emph{Empirical Software Engineering}, vol.~27, no.~6, p. 125, Nov. 2022.

\bibitem{przymusProjectApplicationsGrayscale}
\BIBentryALTinterwordspacing
P.~Przymus, ``Project: {{Applications}} of grayscale data models in software vulnerabilities.'' [Online]. Available: \url{https://radon.nauka.gov.pl/dane/profil/63e5f96bd968d56c86ce5917}
\BIBentrySTDinterwordspacing

\bibitem{piotr_przymus_patchscope_2024}
\BIBentryALTinterwordspacing
{Piotr Przymus}, {Jakub Narębski}, {Mikołaj Fejzer}, and {Krzysztof Stencel}, ``{PatchScope} – {A} {Modular} {Tool} for {Annotating} and {Analyzing} {Contributions},'' 2024, (under review). [Online]. Available: \url{https://ncusi.github.io/PatchScope/articles/patchscope_2024.pdf}
\BIBentrySTDinterwordspacing

\bibitem{DBLP:journals/cacm/WidderG24}
\BIBentryALTinterwordspacing
D.~G. Widder and C.~{Le Goues}, ``What is a 'bug'?'' \emph{Commun. {ACM}}, vol.~67, no.~11, pp. 32--34, 2024. [Online]. Available: \url{https://doi.org/10.1145/3662730}
\BIBentrySTDinterwordspacing

\bibitem{DBLP:conf/sigsoft/WidyasariSLQPTT20}
\BIBentryALTinterwordspacing
R.~Widyasari, S.~Q. Sim, C.~Lok, H.~Qi, J.~Phan, Q.~Tay, C.~Tan, F.~Wee, J.~E. Tan, Y.~Yieh, B.~Goh, F.~Thung, H.~J. Kang, T.~Hoang, D.~Lo, and E.~L. Ouh, ``Bugsinpy: a database of existing bugs in python programs to enable controlled testing and debugging studies,'' in \emph{{ESEC/FSE} '20: 28th {ACM} Joint European Software Engineering Conference and Symposium on the Foundations of Software Engineering, Virtual Event, USA, November 8-13, 2020}, P.~Devanbu, M.~B. Cohen, and T.~Zimmermann, Eds.\hskip 1em plus 0.5em minus 0.4em\relax {ACM}, 2020, pp. 1556--1560. [Online]. Available: \url{https://doi.org/10.1145/3368089.3417943}
\BIBentrySTDinterwordspacing

\bibitem{cve-db}
\BIBentryALTinterwordspacing
T.~M. Corporation. (2023, jan) {CVE} - common vulnerabilities and exposures. [Online]. Available: \url{https://cve.mitre.org/}
\BIBentrySTDinterwordspacing

\bibitem{WelcomePygments}
``Pygments,'' https://pygments.org/, 2023.

\bibitem{Linguist2023}
``{GitHub} {L}inguist,'' https://github.com/github-linguist/linguist, dec 2023.

\bibitem{fleiss1971measuring}
J.~L. Fleiss, ``Measuring nominal scale agreement among many raters.'' \emph{Psychological bulletin}, vol.~76, no.~5, p. 378, 1971.

\bibitem{mills_are_2018}
\BIBentryALTinterwordspacing
C.~Mills, J.~Pantiuchina, E.~Parra, G.~Bavota, and S.~Haiduc, ``Are {Bug} {Reports} {Enough} for {Text} {Retrieval}-{Based} {Bug} {Localization}?'' in \emph{2018 {IEEE} {International} {Conference} on {Software} {Maintenance} and {Evolution} ({ICSME})}, Sep. 2018, pp. 381--392, iSSN: 2576-3148. [Online]. Available: \url{https://ieeexplore.ieee.org/document/8530045}
\BIBentrySTDinterwordspacing

\bibitem{sobreira_dissection_2018}
\BIBentryALTinterwordspacing
V.~Sobreira, T.~Durieux, F.~Madeiral, M.~Monperrus, and M.~de~Almeida~Maia, ``Dissection of a bug dataset: {Anatomy} of 395 patches from {Defects4J},'' in \emph{2018 {IEEE} 25th {International} {Conference} on {Software} {Analysis}, {Evolution} and {Reengineering} ({SANER})}, Mar. 2018, pp. 130--140. [Online]. Available: \url{https://ieeexplore.ieee.org/abstract/document/8330203}
\BIBentrySTDinterwordspacing

\bibitem{DBLP:journals/corr/abs-2312-04730}
\BIBentryALTinterwordspacing
F.~Wu, X.~Liu, and C.~Xiao, ``Deceptprompt: Exploiting llm-driven code generation via adversarial natural language instructions,'' \emph{CoRR}, vol. abs/2312.04730, 2023. [Online]. Available: \url{https://doi.org/10.48550/arXiv.2312.04730}
\BIBentrySTDinterwordspacing

\bibitem{silva2024gitbug}
A.~Silva, N.~Saavedra, and M.~Monperrus, ``Gitbug-java: A reproducible benchmark of recent java bugs,'' \emph{arXiv preprint arXiv:2402.02961}, 2024.

\bibitem{pramod2024bugsphp}
K.~Pramod, W.~De~Silva, W.~Thabrew, R.~Shariffdeen, and S.~Wickramanayake, ``Bugsphp: A dataset for automated program repair in php,'' \emph{arXiv preprint arXiv:2401.07356}, 2024.

\bibitem{DBLP:journals/stvr/GyimesiVSMBFM21}
\BIBentryALTinterwordspacing
P.~Gyimesi, B.~Vancsics, A.~Stocco, D.~Mazinanian, {\'{A}}.~Besz{\'{e}}des, R.~Ferenc, and A.~Mesbah, ``{BUGSJS:} a benchmark and taxonomy of javascript bugs,'' \emph{Softw. Test. Verification Reliab.}, vol.~31, no.~4, 2021. [Online]. Available: \url{https://doi.org/10.1002/stvr.1751}
\BIBentrySTDinterwordspacing

\bibitem{DBLP:conf/msr/KamienskiPBH21}
\BIBentryALTinterwordspacing
A.~V. Kamienski, L.~Palechor, C.~Bezemer, and A.~Hindle, ``Pysstubs: Characterizing single-statement bugs in popular open-source python projects,'' in \emph{18th {IEEE/ACM} International Conference on Mining Software Repositories, {MSR} 2021, Madrid, Spain, May 17-19, 2021}.\hskip 1em plus 0.5em minus 0.4em\relax {IEEE}, 2021, pp. 520--524. [Online]. Available: \url{https://doi.org/10.1109/MSR52588.2021.00066}
\BIBentrySTDinterwordspacing

\bibitem{DBLP:conf/msr/RichterW22}
\BIBentryALTinterwordspacing
C.~Richter and H.~Wehrheim, ``{TSSB-3M:} mining single statement bugs at massive scale,'' in \emph{19th {IEEE/ACM} International Conference on Mining Software Repositories, {MSR} 2022, Pittsburgh, PA, USA, May 23-24, 2022}.\hskip 1em plus 0.5em minus 0.4em\relax {ACM}, 2022, pp. 418--422. [Online]. Available: \url{https://doi.org/10.1145/3524842.3528505}
\BIBentrySTDinterwordspacing

\bibitem{DBLP:conf/msr/MahbubSR23}
\BIBentryALTinterwordspacing
P.~Mahbub, O.~Shuvo, and M.~M. Rahman, ``Defectors: {A} large, diverse python dataset for defect prediction,'' in \emph{20th {IEEE/ACM} International Conference on Mining Software Repositories, {MSR} 2023, Melbourne, Australia, May 15-16, 2023}.\hskip 1em plus 0.5em minus 0.4em\relax {IEEE}, 2023, pp. 393--397. [Online]. Available: \url{https://doi.org/10.1109/MSR59073.2023.00085}
\BIBentrySTDinterwordspacing

\end{thebibliography}

\end{document}